\documentclass[aps,pra,preprint,groupedaddress]{revtex4}

\usepackage{graphicx}
\usepackage{dcolumn}
\usepackage{bm}
\usepackage{color}

\begin{document}


\title{Double diffraction in an atomic gravimeter}

\author{N. Malossi}
\author{Q. Bodart}
\author{S. Merlet}
\author{T. L\'ev\`eque}
\author{A. Landragin}
\author{F. Pereira Dos Santos}

\email[]{franck.pereira@obspm.fr}
\affiliation{LNE-SYRTE, Observatoire de Paris, CNRS,
UPMC, 61 avenue de l'Observatoire, 75014 Paris, FRANCE}

\date{\today}

\begin{abstract}

We demonstrate the realization of a new scheme for cold atom
gravimetry based on the use of double diffraction beamsplitters
recently demonstrated in \cite{Leveque}, where the use of two
retro-reflected Raman beams allows symmetric diffraction in $\pm
\hbar k_{eff}$ momenta. Though in principle restricted to the case
of zero Doppler shift, for which the two pairs of Raman beams are
simultaneously resonant, we demonstrate that such diffraction
pulses can remain efficient on atoms with non zero velocity, such
as in a gravimeter, when modulating the frequency of one of the
two Raman laser sources. We use such pulses to realize an
interferometer insensitive to laser phase noise and some of the dominant systematics.
This reduces the technical requirements and would allow the realization of a simple atomic gravimeter.
We demonstrate a sensitivity of $1.2\times10^{-7}g$ per shot.

\end{abstract}


\maketitle

\section{Introduction}

In the last decade atom interferometry has been increasingly
applied to the inertial sensor domain so that state of the
art atom gravimeters have reached sensibilities comparable to
the commercial ones \cite{Peters01,LeGouet08}. The number of possible applications of
gravimetry, ranging from fundamental physics \cite{Fixler07,Lamporesi08},
metrology \cite{Geneves,Merlet08} and geophysics to industrial related
applications, such as navigation and underground prospection ..., have
stimulated the research towards higher sensitivity, stability and
overall performance of atomic inertial sensors and towards more
compact and movable systems. The present experimental set-up, exploiting an atomic interferometer with Raman
transitions \cite{Kasevich91}, has been originally developed as a prototype for the
French Watt-Balance experiment\cite{Geneves}, aiming thus to both high
accuracy and discrete portability (interferometer
interaction length of a few centimeters only) which, on the other hand, reduces the
sensitivity of the interferometer by limiting the interaction time.\\
A possible solution for increasing the sensitivity of the
interferometers is to increase the separation between the
atomic wave packets leaving the first beamsplitter of the
interferometer, enlarging thus the arms separation. Material
gratings \cite{Ekstrom95}, magneto-optical beam-splitters \cite{Pfau93,Schumm05},
momentum transfer by adiabatic passage \cite{Weitz94}, Kapitza-Dirac \cite{Cahn97} and Bragg diffraction
\cite{Rasel95,Giltner95} with recent development \cite{Muller08}, Bloch oscillations
\cite{Clade09,Muller09} and finally Doppler-free double diffraction in Raman
configuration \cite{Leveque} have
been implemented for obtaining coherently larger
separation angle of the atomic wavepackets.

In this paper we show how the double diffraction scheme described in \cite{Leveque} can be extended to the case of an atomic gravimeter, despite the increasing Doppler shift due to gravity. This scheme consists
essentially in the transfer of the same amount of photon momentum $2\hbar k$ in opposite directions
from the Raman lasers to both arms of the
interferometer, where $k$ is the wavevector of the Raman lasers. This doubles the separation between the atomic wavepackets with respect to the usual configuration of an interferometer based on Raman transitions \cite{Kasevich91} where only one arm gains the momentum transfer. It thus allows increasing the intrinsic sensitivity of the interferometer, and thus the sensitivity to g when the interferometer phase noise is limited by the detection noise or the electronic phase noise on the Raman laser phase difference. When the interferometer noise is dominated by parasitic vibrations, no gain in sensitivity to g is expected. Another advantage of this scheme is to relax the requirements on subsystems, such as the phase noise of the microwave reference or the efficiency of magnetic shielding.

\section{Double diffraction}

The Raman transitions couples the two hyperfine ground states of
$^{87}$Rb atoms (named $|g\rangle$ and $|e\rangle$) by using two
lasers with frequencies (labelled  $\omega_1$ and  $\omega_2$)
which are detuned to the red of the $D_{2}$ line so
that the frequency difference between the lasers matches the
hyperfine transition frequency.

In our experiment, the two Raman beams are overlapped on a
polarizing beam splitter with orthogonal linear polarizations and
are injected in the same polarization maintaining fiber to be sent
to the vacuum chamber.\\
At the output of the fiber, the beams pass
through a quarter-wave plate which turn the two polarizations into
opposite circular ones. These beams are then finally retro-reflected after passing through a second
quarter-wave plate.
For the two-photon transitions $|F=2,m_F=0\rangle\rightarrow|F=1,m_F=0\rangle$ there are thus
two pairs of beams (with polarizations $\sigma^{+}/\sigma^{+}$ and
$\sigma^{-}/\sigma^{-}$) which can drive the counter-propagating
Raman transitions with effective wave-vectors
$\pm\mathbf{k}_\mathrm{eff} = \pm(\mathbf{k}_1-\mathbf{k}_2)$. The
ground state $|g,\mathbf{p}\rangle$ is thus coupled to
$|e,\mathbf{p}+\hbar \mathbf{k}_\mathrm{eff}\rangle$ by one of the
pairs and to $|e,\mathbf{p}-\hbar \mathbf{k}_\mathrm{eff}\rangle$
by the other pair.

In the absence of Doppler shift both pairs are  simultaneously resonant so that
the atomic population is diffracted into two states
$|e,\mathbf{p}\pm\hbar \mathbf{k}_\mathrm{eff}\rangle$. This
degeneracy was recently used to realize an interferometer where
the difference between the momenta in the two arms is $2\hbar \mathbf{k}_\mathrm{eff}$.
This scheme increases the inertial sensitivity by a factor 2 and
can be combined with extra Raman pulses to further increase the
area of the interferometer \cite{Leveque}.

In this article, on the other hand, we generalize the previous
scheme to atoms with non zero Doppler shift, allowing the
realization of an interferometer for gravity measurements. In order
to ensure simultaneous resonance with the two transitions while the atoms are free falling, a third frequency is required in the Raman
beams. Keeping a two laser beams geometry, the three different
frequencies can be gained by modulating one of the two lasers, to
generate sidebands with adjustable separation. Sweeping the
modulation frequency allows to fulfill the resonance condition of
the falling atoms.

\begin{figure}[!h]
\includegraphics[width=5cm,angle=270]{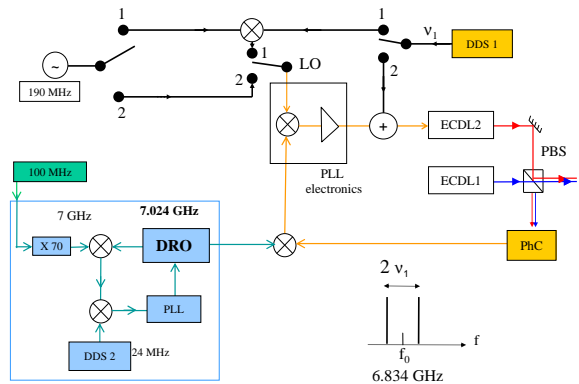}
 \caption{Scheme of the phase lock loop (PLL). In case 1, the modulation is carried by the Local Oscillator (LO).
 In case 2, it is directly applied onto the current of the extended cavity diode laser (ECDL). DRO : Dielectric Resonator Oscillator, PhC : Photoconductor.}
 \label{pll}
 \end{figure}

Figure \ref{pll} displays a scheme of the phase lock loop (PLL) used to servo the phase difference between the lasers. Modulation of the laser was realized with two
different techniques: using a modulated local oscillator (LO), or by adding a
modulation on the current of the diode laser. In the first case,
the modulated LO is generated by mixing a RF synthesizer at 190 MHz
with a Direct Digital Synthesizer (DDS1) whose frequency $\nu_1$ matches the Doppler shift. The LO then contains two frequencies: $190\pm\nu_1$ MHz. The PLL
can efficiently copy the modulated signal onto the frequency
difference between the lasers as long as the modulation frequency
is smaller than the bandwidth of the PLL (about 4 MHz).\\
In the second case, the modulation is obtained by directly modulating the current of the diode. Since the PLL
effectively works against this modulation, but with a finite gain, an
efficient frequency modulation can still be obtained by adjusting
the amplitude of the modulation signal. In practice, the
modulation amplitude is set to minimize the carrier (more than 40
dB rejection) in order to reduce the influence of parasitic
copropagating transitions. Such transitions, though in principle forbidden by polarization selection rules, would still occur
due to the relatively small detuning of the carrier (shifted by the recoil frequency $\omega_R$) and imperfections in the polarizations of the Raman beams.\\
In the standard configuration, the frequency difference between
the lasers can then be adjusted at will using either DDS1, which
is then used directly as the LO for the phase comparison, or DDS2. In the
cases studied here, modifying DDS2 changes the average frequency
difference while changing DDS1 modifies the spacing between the
sidebands.

\subsection{Experimental setup}
The experiment was carried out in the gravimeter setup described in detail in~\cite{GraviAPB1,LeGouet08}. In
this compact experimental set-up, cold $^{87}Rb$ atoms are first
trapped in a 3D MOT during 300 ms. With respect to \cite{LeGouet08}
where a 2D MOT was used, atoms are loaded directly from a
background Rb vapor. Atoms are further cooled during a brief
optical molasses phase before being released by switching off the
cooling lasers. A sequence of microwave and optical pulses is then
used to select atoms in the $|F=2,m_F=0\rangle$ state. A selection pulse
using a standard Raman pulse (diffraction in only one direction)
is applied during this sequence in order to reduce the width
of the velocity distribution before the atoms experience the
interferometer. During their free fall over a few centimeters, the
interferometer is obtained by pulsing counterpropagating Raman lasers in the
vertical direction. A three pulse sequence allows splitting, deflecting and recombining the atomic wavepackets.
These three successive Raman pulses are
separated by the free evolution times of up to $T=50$ ms. After
the interferometer, the populations in the two hyperfine states are
measured using a fluorescence detection technique
\cite{LeGouet08}.


\subsection{Spectroscopy and Rabi oscillations}

\begin{figure}[!h]
 \includegraphics[width=8.5cm]{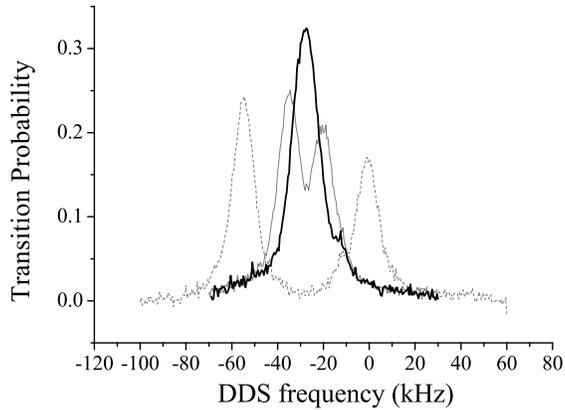}
 \caption{Raman resonance spectra, obtained by scanning the average Raman frequency DDS2, across the resonance, for three different modulations frequencies: 350 kHz (dashed line), 370 kHz (thin line) and 375 kHz (thick line). At 375 kHz, a single peak is observed indicating that the two resonance conditions for diffracting up and down are satisfied.}
 \label{spectro}
 \end{figure}

As a first step of our investigation, we realize the spectroscopy
of the Raman transition. For this measurement, the duration of the selection pulse is set to 120 $\mu$s.
After the selection, we apply a relatively long single "double diffraction" Raman pulse of 110 $\mu$s
17 ms after releasing the atoms from the molasses. The Raman laser
intensities are adjusted in order to maximize the transfer efficiency at
resonance. Figure \ref{spectro} displays the measured transition
probability as a function of the frequency of DDS2 for three
different modulations frequencies: 350, 370, 375 kHz. At 350 and
370 kHz, we observe two peaks which correspond to individual
resonance with one of the two sidebands. For this measurement, the modulation was applied on the diode laser current, which typically results in asymmetric sidebands, as it is evident from the data. This asymmetry results from the cumulative effects of phase and amplitude modulation. At 375 kHz, these two
peaks merge, indicating that the two sidebands are simultaneously
resonant. Therefore they are able to diffract atoms in the states
$|p\pm \hbar k_{eff}\rangle$ at the same time.

\begin{figure}[!h]
 \includegraphics[width=12cm]{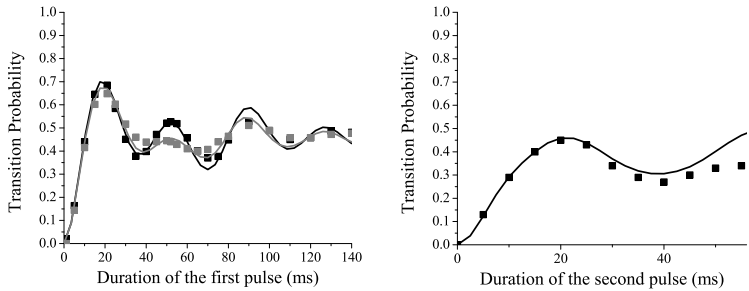}
 \caption{Transfer efficiency of a double diffraction pulse. Squares are measured values and lines are the results of the numerical simulation. Left: first pulse,
 for atoms selected with standard Raman pulses of durations 23 $\mu$s (grey squares and line) and 60 $\mu$s (black squares and line).  Right: second pulse after a first pulse of duration 18 $\mu$s and a pusher beam.}
 \label{rabi1}
 \end{figure}

Having then selected the most efficient modulation
frequency, 375 kHz, we have measured, for atoms that have been velocity selected, the transition probability as a
function of the Raman pulse duration at the maximum power
available. Figure \ref{rabi1} displays at the left the
measurements of the transition probability for two different
pre-selecting Raman transitions of duration 23 and 60 $\mu$s (grey and black squares respectively).
A maximum transfer efficiency of about 70 \% is obtained for a
first pulse of duration 18 $\mu$s.

After a first pulse of duration 18 $\mu$s, atoms left in $|F=2\rangle$
are cleared by a pulse of a pusher beam which is tuned on the
$|F=2\rangle\rightarrow |F'=3\rangle$ transition. A second Raman pulse is then
performed whose transfer efficiency versus duration is plotted in
figure \ref{rabi1} at the right.

To discuss the physics of the double diffraction process, we will consider for simplicity the evolution of the
quantum state within a three states basis, $|g,\mathbf{p}\rangle$,
$|e,\mathbf{p}+\hbar \mathbf{k}_\mathrm{eff}\rangle$, $|e,\mathbf{p}-\hbar \mathbf{k}_\mathrm{eff}\rangle$. If the initial state is $|g,\mathbf{p}\rangle$, the transition probability undergoes Rabi oscillations between the two states $|g,\mathbf{p}\rangle$ and $|e,\mathbf{p}+\hbar \mathbf{k}_\mathrm{eff}\rangle + |e,\mathbf{p}-\hbar \mathbf{k}_\mathrm{eff}\rangle$, with a Rabi frequency $\Omega_{dd}$. A 18 $\mu$s long pulse thus corresponds for our experimental parameters to a $\pi$-pulse of this oscillation. If now the initial state is $|e,\mathbf{p}+\hbar \mathbf{k}_\mathrm{eff}\rangle$, there is a Rabi oscillation between the two states $|e,\mathbf{p}+\hbar \mathbf{k}_\mathrm{eff}\rangle$ and $|e,\mathbf{p}-\hbar \mathbf{k}_\mathrm{eff}\rangle$, at the frequency $\Omega_{dd}/2$. With respect to this second Rabi
oscillation, the second pulse of the interferometer is a $\pi$ pulse, when it is a $2\pi$ pulse with respect the first Rabi oscillation.

In practice, the evolutions of the transition probability versus pulse duration are found to differ
significantly from the expected sinusoidal dependance of the Rabi oscillations. In particular, the limited efficiency of the second pulse will reduce the contrast of the interferometer as it will be evident from the data. This deviation from the sinusoidal behaviour arises from averaging over the initial vertical velocity distribution (the discussion above holds only for plane waves perfectly resonant), couplings to higher momenta states and spontaneous emission. We performed the numerical calculation of the evolution of the transition probability, taking into account these effects, and extending the basis to the five lowest coupled momenta states. The results of this numerical model are displayed as lines on figure \ref{rabi1} for comparison with the experimental data points. The calculations explain fairly well the observed behaviours. In particular, we find that the averaging over the initial velocity is the dominant contribution to the imperfection of the double diffraction pulses.

\section{The interferometer}

\begin{figure}[!h]
 \includegraphics[width=5cm,angle=270]{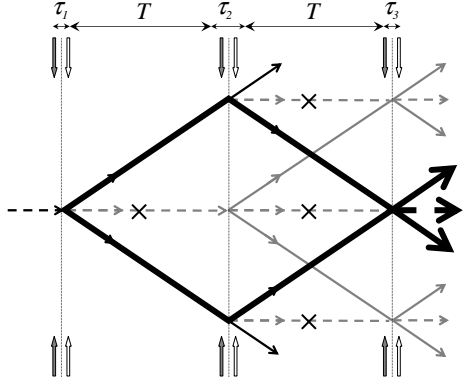}
 \caption{Scheme of the interferometer, realized with a sequence of three double diffraction pulses of durations $\tau_i$, separated by free evolutions of duration $T$. Little arrows represent the retroreflected Raman lasers. Thick lines represent the two interfering paths that create the double diffraction interferometer. Thin lines represent parasitic paths, which eventually interfere to create parasitic interferometers. Crosses have been put on the paths removed by pusher beam pulses. Solid (resp. dashed) lines represent partial wavepackets in state $|e\rangle$ (resp. $|g\rangle$).}
 \label{interfero}
 \end{figure}

A sequence of three double diffraction Raman transitions finally allows realizing a double diffraction interferometer, whose geometry is displayed in figure \ref{interfero}.
After the first (and eventually the second
pulse as well), a pulse of pusher beam, resonant with the $|F=2\rangle\rightarrow |F'=3\rangle$, clears the atoms
in $|F=2\rangle$. This prevents forming parasitic interferometers with the
atoms initially left in $|F=2\rangle$.

\begin{figure}[!h]
 \includegraphics[width=8.5cm]{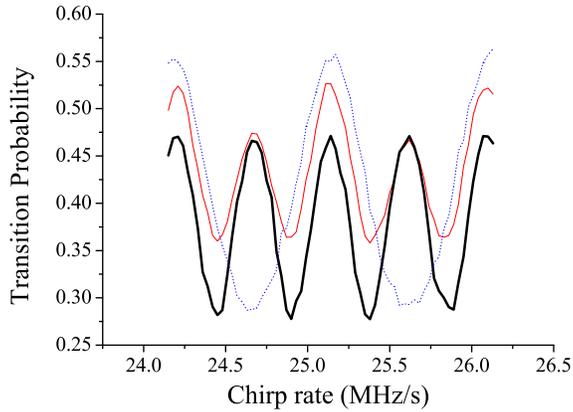}
 \caption{Fringe patterns obtained when modulation the local oscillator for $2T=2$ ms: double diffraction interferometer with pulse sequence 18-36-18 $\mu$s (solid thick line), same pulse sequence without pusher pulse (thin solid line), Raman pulses of half durations (dashed line)}
 \label{dd1ms}
 \end{figure}

The phase of the interferometer can be scanned by changing the chirp rate $a$ applied to DDS1 in order to compensate for the increasing Doppler shift, and thus to keep the Raman lasers on resonance. Figure \ref{dd1ms} displays several
interferometer fringe patterns obtained by scanning the chirp rate
$a$, depending on whether a clearing pulse is applied. For this
data set, sidebands are produced using a modulated LO and the
total interferometer time is set to $2T=2$ ms. The thick solid line
corresponds to the case of a three pulses interferometer (of
durations 18-36-18 $\mu$s), with a pulse of pusher beam right
after the first pulse. As expected for a double diffraction
interferometer, the fringe spacing scales as $2aT^2$. We also
checked that the phase of this double diffraction interferometer
is insensitive to a phase jump of the Raman laser phase
difference. The thin solid line corresponds to the same Raman
lasers pulse sequence, except that the pusher pulse is not used.
The fringe pattern exhibits beats between two different fringe
spacings, which scale respectively as $aT^2$ and $2aT^2$. If
reducing the pulse durations by a factor of 2, parasitic
interferometers of half area (shown with dashed lines on figure
\ref{dd1ms}) dominate the fringe pattern.

\begin{figure}[!h]
 \includegraphics[width=8.5cm]{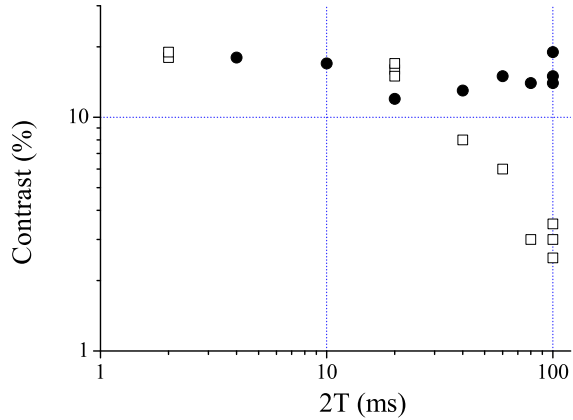}
 \caption{Evolution of the contrast of the interferometer versus total interferometer
 duration $2T$ for two different techniques: modulated local oscillator (open squares) and modulation of the laser current (full circles).}
 \label{cont}
 \end{figure}

The contrast of the interferometer is measured for the
two modulation techniques as a function of the interferometer duration $2T$. The results, obtained using two pusher pulses, are displayed on figure
\ref{cont}. The contrast is defined here as the difference between the minimum and the maximum of the transition probability.
When expressed in \%, it is this difference multiplied by 100.
We find a rapid decrease of the contrast with
interaction time for the modulated LO case, which is due to
the finite bandwidth of the PLL, which is about 4 MHz. This decay
would be less pronounced if increasing the bandwidth of the PLL,
by using for instance a laser with an intracavity EOM \cite{LeGouet09}. In
the case where sidebands are produced by modulating the current,
the contrast remains almost constant, on the order of 20 \%.
In the latter case, the contrast was found to depend strongly on
the operating point of the laser and especially on the tuning
sensitivity versus current, which depends on the proximity of mode
hops in the laser frequency.

\begin{figure}[!h]
 \includegraphics[width=8.5cm]{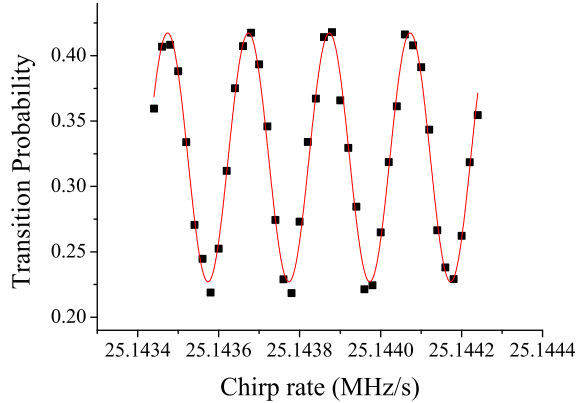}
 \caption{Double diffraction interferometer fringe pattern, obtained using the current modulation technique, for a total interferometer duration of $2T=100$ ms}
 \label{contopt}
 \end{figure}

Figure \ref{contopt} shows an interferometer pattern with $2T=100$
ms and $\tau=18\mu$s. The contrast is 19\%. The number of detected atoms is $2\times10^5$. The signal to noise
ratio measured at mid-fringe is $1/\sigma_{Phi}=10$, which
corresponds to a sensitivity of $1.2\times10^{-7}$g per shot.
Influence of residual vibrations is reduced thanks to the
combination of a passive isolation platform, and a post-correction
scheme using the independent measurement of vibrations with a low
noise sismometer \cite{LeGouet08}. Averaging such a fringe pattern
for about an hour allows for searching with a high resolution the
eventual contribution of single area parasitic interferometer.
Removing the result of a sinusoidal fit with $2aT^2$ scaling
reveals well resolved residuals of $aT^2$ scaling, whose amplitude
is about 1\% of the amplitude of the fringes. We believe such
parasitic contributions arise from residual
copropagating transitions, due to imperfect extinction of the carrier.

\section{Systematics}
\begin{figure}[!h]
 \includegraphics[width=8.5cm]{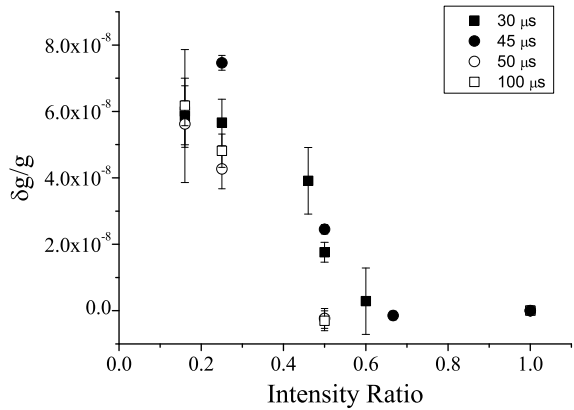}
 \caption{Relative shift of the interferometer phase as a function of the Raman beam intensity. The intensity is normalized to the maximum intensity available,
 for which maximum transfer efficiency of a single pulse corresponds to a duration of $18\mu$s. The different symbols correspond to different durations of the selection pulse.}
 \label{inflpuiss}
 \end{figure}

In principle, the phase of the interferometer should be insensitive to magnetic field gradients and temporal fluctuations and
AC Stark shifts, as the two partial
wavepackets propagate in the same internal state in the two arms
of the interferometer. Measurements of the average interferometer
phase as a function of the Raman beam intensities are displayed in
figure \ref{inflpuiss}. The duration of the Raman pulses are adjusted in order to keep the product $\Omega\tau$ constant.
 Well resolved shifts are observed, whose scaling
with power does not follow a simple linear dependence. The observed
behavior was found to be robust versus changes in the width of the
selected velocity class. These phase shifts might be due to the
presence of higher order sidebands, which shift the Raman
resonance conditions. The detailed analysis of this effect, beyond the scope of this paper, would deserve further investigations.

\section{Conclusion}
We extend the interferometer scheme based on double diffraction of
\cite{Leveque} to the case of a vertical accelerometer.
This interferometer geometry results in the increase by a factor 2 of
the scale factor, and a reduced sensitivity to phase noise and systematics with respect to the traditional geometry.
Two different methods are studied, using either a double frequency
Local Oscillator in the PLL or a modulation of the frequency of
one of the two Raman lasers. Best results are obtained with the latter technique, in which the two lowest order
sidebands are used to ensure simultaneous resonance of Raman
transitions with effective wave vectors in the up or down
directions. It allows reaching a good contrast of 19 \% for an
interferometer of duration $2T=100$ ms, and a sensitivity of $1.2\times10^{-7}$g per shot for a repetition rate of about 2 Hz.
The disadvantages of this last method are parasitic transitions
induced by the carrier frequency and light shifts induced by the
higher order sidebands. These drawbacks could be avoided for instance by using
three independent lasers to generate the three required
frequencies. Alternatively, the required frequencies could be obtained by phase modulating with an EOM a single laser, using a two frequency microwave reference signal.
Finally, further increase of the scale factor can easily be realized with multi-$\hbar k$ beam splitters \cite{Leveque}.
Such a geometry, with increased sensitivity, could allow developing more compact sensors, with reduced complexity, as important
constraints in the design of the experiment are relaxed by the insensitivity to magnetic field gradients and Raman laser phase noise.

\begin{acknowledgments}
We thank the Institut Francilien pour la Recherche sur les Atomes
Froids (IFRAF), the European Union (FINAQS contract) and the European Science Foundation (EUROQUASAR projet) for financial
supports. Q. B. thanks CNES for supporting his work.
\end{acknowledgments}



\begin{thebibliography}{30}
\bibitem{Leveque} T. L\'ev\`eque, A. Gauguet, F. Michaud, F. Pereira Dos Santos, and A. Landragin, Phys. Rev. Lett. {\bf 103}, 080405 (2009)

\bibitem{Peters01} A. Peters, K. Y. Chung, S. Chu, Metrologia \textbf{38}, 25 (2001)

\bibitem{LeGouet08} J. Le Gou\"{e}t, T. E. Mehlst\"{a}ubler, J. Kim, S. Merlet, A. Clairon, A.
Landragin, F. Pereira Dos Santos, Appl. Phys B {\bf 92}, 133 (2008)
\bibitem{Fixler07} J. B. Fixler, G. T. Foster, J. M. McGuirk, M. A. Kasevich, Science Magazine Vol. 315. no. 5808, 74 (2007).
\bibitem{Lamporesi08} G. Lamporesi, A. Bertoldi, L. Cacciapuoti, M. Prevedelli, and G. M. Tino, Phys. Rev. Lett. \textbf{100}, 050801 (2008)
\bibitem{Geneves} G. Genev{\`e}s, P. Gournay, A. Gosset, M. Lecollinet, F. Villar, P. Pinot,
P. Juncar, A. Clairon, A. Landragin, D. Holleville, F. Pereira Dos
Santos, J. David, M. Besbes, F. Alves, L. Chassagne, S. Top\c{c}u, IEEE Transactions on
Instrumentation and Measurement {\bf 54}, 850-853 (2005)

\bibitem{Merlet08} S. Merlet, A. Kopaev, M. Diament, G. Geneves, A. Landragin and F. Pereira Dos Santos, Metrologia \textbf{45}, 265-274 (2008)

\bibitem{Kasevich91} M. Kasevich and S. Chu, Phys. Rev. Lett. \textbf{67}, 181 (1991)


\bibitem{Ekstrom95} C. Ekstrom, J. Schmiedmayer, M. Chapman, T. Hammond and D. E. Pritchard, Phys. Rev. A \textbf{51}, 3883 (1995)

\bibitem{Pfau93} T. Pfau, Ch. Kurtsiefer, C.S. Adams, M. Sigel, and J.
Mlynek, Phys. Rev. Lett. {\bf 71}, 3427 (1993)
\bibitem{Schumm05} T. Schumm et al., Nat. Phys. {\bf 1}, 57 (2005)
\bibitem{Weitz94} M. Weitz, B.C. Young, and S. Chu, Phys. Rev. Lett. {\bf 73}, 2563 (1994).

\bibitem{Cahn97} S. B. Cahn, A. Kumarakrishnan, U. Shim, T. Sleator, P. R. Berman and B. Dubetsky, Phys. Rev. Lett. {\bf 79}, 784 (1997)



\bibitem{Rasel95} E.M. Rasel, M.K. Oberthaler, H. Batelaan, J. Schmiedmayer,
and A. Zeilinger, Phys. Rev. Lett. {\bf 75}, 2633 (1995)
\bibitem{Giltner95} D.M. Giltner, R.W. McGowan, and S.A. Lee, Phys. Rev.
Lett. {\bf 75}, 2638 (1995)

\bibitem{Muller08} H. M\"{u}ller, S.-W. Chiow, Q. Long, S. Herrmann, and S.
Chu, Phys. Rev. Lett. {\bf 100}, 180405 (2008)

\bibitem{Clade09} P. Clad\'e, S. Guellati-Kh\'elifa, F. Nez, and F. Biraben, Phys. Rev. Lett. \textbf{102}, 240402 (2009)

\bibitem{Muller09} H. M\"{u}ller, S.-W. Chiow, Q. Long, S. Herrmann, and S.
Chu, Phys. Rev. Lett. 102, 240403 (2009)




\bibitem{GraviAPB1} P. Cheinet, F. Pereira Dos Santos , T. Petelski, J. Le Gou{\"e}t, J. Kim,
K.T. Therkildsen, A. Clairon and A. Landragin, Appl. Phys. B {\bf
84}, 643 (2006)

\bibitem{LeGouet09} J. Le Gou\"{e}t, J. Kim, C. Bourassin-Bouchet, M. Lours, A. Landragin, F. Pereira Dos Santos, Optics Comm. {\bf 282}, 977-980 (2009)




\end{thebibliography}
\end{document}